\documentclass[11pt]{article}
\usepackage{moriond,epsfig}

\def\D0{D\O }

\def\be{\begin{equation}}
\def\ee{\end{equation}}
\def\bea{\begin{eqnarray}}
\def\eea{\end{eqnarray}}

\begin{document}

%------------------------BEGINNING OF NIKHEF TITLE PAGE----------------------

\thispagestyle{empty}

\begin{flushright}
Nikhef-2011-017\\
\end{flushright}

\vspace{3.0truecm}
\begin{center}
\boldmath
\large\bf In Pursuit of New Physics with B Decays
\unboldmath
\end{center}

\vspace{0.9truecm}
\begin{center}
Robert Fleischer\\[0.1cm]
{\sl Nikhef, Science Park 105, 
NL-1098 XG Amsterdam, The Netherlands}
\end{center}

\vspace{2.9truecm}

\begin{center}
{\bf Abstract}
\end{center}

{\small
\vspace{0.2cm}\noindent
Decays of $B$ mesons offer interesting probes to search for physics beyond 
the Standard Model. Thanks to the data taking at the LHC, we are at the beginning of a new era 
of precision $B$ physics. I will discuss recent developments concerning the analyses of promising 
channels to search for signals of New Physics at the LHC, $B^0_s\to J/\psi \phi$, 
$B^0_s\to K^+K^-$ and $B^0_s\to\mu^+\mu^-$.}

\vspace{3.9truecm}

\begin{center}
{\sl Invited talk at Rencontres de Moriond 2011, QCD and High Energy Interactions\\
La Thuile, Italy, 20--27 March 2011\\
To appear in the Proceedings}
\end{center}

\vfill
\noindent
May  2011

\newpage
\thispagestyle{empty}
\vbox{}
\newpage
 
\setcounter{page}{1}

%------------------------END OF NIKHEF TITLE PAGE------------------------------

\vspace*{4cm}
\title{IN PURSUIT OF NEW PHYSICS WITH B DECAYS}

\author{ROBERT FLEISCHER}

\address{Nikhef, Theory Group, 
Science Park 105\\
NL-1098 XG Amsterdam, The Netherlands}

\maketitle\abstracts{Decays of $B$ mesons offer interesting probes to search for physics beyond 
the Standard Model. Thanks to the data taking at the LHC, we are at the beginning of a new era 
of precision $B$ physics. I will discuss recent developments concerning the analyses of promising 
channels to search for signals of New Physics at the LHC, $B^0_s\to J/\psi \phi$, 
$B^0_s\to K^+K^-$ and $B^0_s\to\mu^+\mu^-$.}

\section{Setting the Stage}
The lessons from the data on weak decays of $B$, $D$ and $K$ mesons collected so far 
is that the Cabibbo--Kobayashi--Maskawa (CKM) matrix is the dominant source of flavour 
and CP violation. New effects have not yet been established although there are potential 
signals in the flavour sector. The implications for the structure of New Physics (NP) is
that we may actually have to deal with a large characteristic NP scale $\Lambda_{\rm NP}$, 
i.e.\ one that is not just in the TeV regime, or (and?) that symmetries prevent large NP effects in  
the flavour sector. 

The best known example of the latter feature is ``Minimal Flavour Violation" (MFV), 
where -- sloppily speaking -- CP and flavour violation is essentially the same as in 
the Standard Model (SM). However, it should be emphasized that MFV is still far from being 
experimentally established and that there are various non-MFV scenarios with room for sizeable 
effects.\cite{buras} 

Nevertheless, we have to be prepared to deal with smallish NP signals. But the excellent news is 
that we are at the beginning of a new era in particle physics, the LHC era, which will also bring us 
to new frontiers in high-precision $B$ physics.

\section{Promising B-Physics Probes to Search for New Physics}
\subsection{$B^0_s\to J/\psi \phi$}
A particularly interesting decay is $B^0_s\to J/\psi \phi$, offering a sensitive probe for
CP-violating NP contributions to $B^0_s$--$\bar B^0_s$ mixing. In the SM, the corresponding
CP-violating phase $\phi_s$ is tiny, taking a value of about $-2^\circ$. However, NP effects
may well generate a sizable phase, which actually happens in various specific extensions
of the SM.\cite{buras} This phase enters the mixing-induced CP violation 
${\cal A}_{\rm CP}^{\rm mix}$ in $B^0_s\to J/\psi \phi$, which 
arises from the interference between $B^0_s$--$\bar B^0_s$ mixing and decay processes. 
As the final state is a mixture of CP-odd and CP-even eigenstates, a time-dependent angular 
analysis of the $J/\psi [\to\mu^+\mu^-]\phi [\to\ K^+K^-]$ decay products is needed in order to
disentangle them.\cite{DDF} Neglecting doubly Cabibbo-suppressed terms of the decay 
amplitude (see below), hadronic parameters cancel and 
${\cal A}_{\rm CP}^{\rm mix}=\sin\phi_s$. Since the tiny SM value of $\phi_s$ implies 
smallish CP violation in $B^0_s\to J/\psi \phi$, this channel plays a key role in the
search for NP.\cite{DFN}

Since a couple of years, measurements of CP violation in $B^0_s\to J/\psi \phi$ at
the Tevatron indicate possible NP effects in $B^0_s$--$\bar B^0_s$ mixing,
which are also complemented by the measurement of the anomalous like-sign dimuon
charge asymmetry at \D0. LHCb has also joined the arena, with a first untagged analysis
reported at Moriond 2011,\cite{LHCb-phis} and a first tagged analysis reported a couple of 
weeks later. Interestingly, this measurement points to a picture similar to that obtained at the 
Tevatron, with a sign of a sizable, negative value of $\phi_s$. However, the large uncertainties 
preclude definite conclusions. Fortunately, the prospects for analyses of CP 
violation in $B^0_s\to J/\psi \phi$ are excellent: the 2011 LHCb data 
($1~\mbox{fb}^{-1}$) should allow the world's best measurement of $\phi_s$. 

SM penguin effects, which are described by a CP-conserving hadronic parameter $ae^{i\theta}$, 
are usually neglected in these analyses. They enter the decay amplitude in the following form:
\begin{equation}
A(B^0_s\to J/\psi \phi)\propto {\cal A}\left[1+\epsilon (a e^{i\theta}){e^{i\gamma}} \right], 
\end{equation}
where $\epsilon\equiv\lambda^2/(1-\lambda^2)=0.05$ with $\lambda\equiv|V_{us}|$,
and modify the mixing-induced 
CP asymmetry as ${\cal A}_{\rm CP}^{\rm mix}\propto\sin(\phi_s+\Delta\phi_s)$.\cite{FFM}
The hadronic shift $\Delta\phi_s$ can be controlled through $B^0_s\to J/\psi \bar K^{*0}$, 
which has recently been observed by the CDF and LHCb collaborations. Two scenarios 
emerge: 
\begin{itemize}
\item {\it Optimistic:} large ${\cal A}_{\rm CP}^{\rm mix}\sim-40\%$ would be an 
unambiguous signal of NP.
\item {\it Pessimistic:} smallish ${\cal A}_{\rm CP}^{\rm mix}\sim-(5...10)\%$ would require 
more work  to clarify the picture.
\end{itemize}
The hadronic shift $\Delta\phi_s$ must in any case be controlled in order to match the future 
LHCb experimental precision, in particular for an LHCb upgrade.

An interesting probe for similar penguin topologies is also provided by the
$B^0_s\to J/\psi K_{\rm S}$ channel,\cite{RF-BspsiKS} which is related to 
$B^0_d\to J/\psi K_{\rm S}$ by the $U$-spin symmetry of strong interactions, allowing 
a determination of the angle $\gamma$ of the unitarity triangle and the control of the penguin 
uncertainties in the extraction of the $B^0_d$--$\bar B^0_d$ mixing phase $\phi_d$ from 
$B^0_d\to J/\psi K_{\rm S}$. CDF has recently observed the $B^0_s\to J/\psi K_{\rm S}$ mode. 
A detailed phenomenological analysis has recently been performed,\cite{DeBFK} with a first 
(toy) feasibility study for LHCb. This study has shown that 
the determination of $\gamma$ is feasible, while the main application will be the control of the 
penguin effects. The $B^0_s\to J/\psi K_{\rm S}$ decay looks particularly interesting for the 
LHCb upgrade. 

\subsection{$B^0_s\to K^+K^-$}
The decay $B^0_s\to K^+K^-$ can be related to $B^0_d\to\pi^+\pi^-$ by means of the $U$-spin
symmetry of strong interactions.\cite{RF-BsK+K-} In the SM, the decay amplitudes can be 
written as follows:
\begin{equation}\label{ampl}
A(B_s^0\to K^+K^-)\propto {\cal C}'\left[ e^{i\gamma}+d'e^{i\theta'}/\epsilon\right], 
\quad
A(B_d^0\to\pi^+\pi^-)\propto {\cal C}\left[ e^{i\gamma}-d\,e^{i\theta}\right],
\end{equation}
where ${\cal C}$,  ${\cal C}'$ and $d\,e^{i\theta}$,  $d'\,e^{i\theta'}$ are CP-conserving
``hadronic" parameters. The $U$-spin symmetry implies $d'=d$, $\theta'=\theta$, allowing the determination both of $\gamma$ and of the hadronic parameters $d(=d')$,
$\theta$ and $\theta'$ from the observables of the $B^0_s\to K^+K^-$, $B^0_d\to\pi^+\pi^-$ 
system.\cite{RF-BsK+K-} At LHCb, thanks to precise measurements of the corresponding 
CP-violating asymmetries, this strategy is expected to give an experimental accuracy for 
$\gamma$ of only a few degrees.\cite{LHCb-roadmap}

In order to get ready for the LHCb data (and improved Tevatron measurements), an
analysis of  $B^0_s\to K^+K^-$ was recently performed.\cite{FK} As input, it uses
$B$-factory data in combination with BR$(B_s\to K^+K^-)$ measurements by CDF and Belle at the 
$\Upsilon(5S)$ resonance as well as updated information on $U$-spin-breaking form-factor 
ratios. The following result for $\gamma$ is obtained:
\begin{equation}
\gamma=(68.3^{+4.9}_{-5.7}|_{\rm input}
        \mbox{}^{+5.0}_{-3.7}|_\xi\mbox{}^{+0.1}_{-0.2}|_{\Delta\theta})^\circ,
\end{equation}
where $ \xi\equiv d'/d = 1 \pm 0.15$ and $\Delta\theta\equiv\theta'-\theta = \pm 20^\circ$
was assumed to explore the sensitivity to $U$-spin-breaking effects. This result is in 
excellent and remarkable agreement with the fits of the unitarity triangle, 
$\gamma=(67.2\pm3.9)^\circ$ [CKMfitter] and $(69.6\pm3.1)^\circ$ [UTfit].

\begin{figure}
\centerline{
\begin{tabular}{cc}
  \includegraphics[width=7.0truecm]{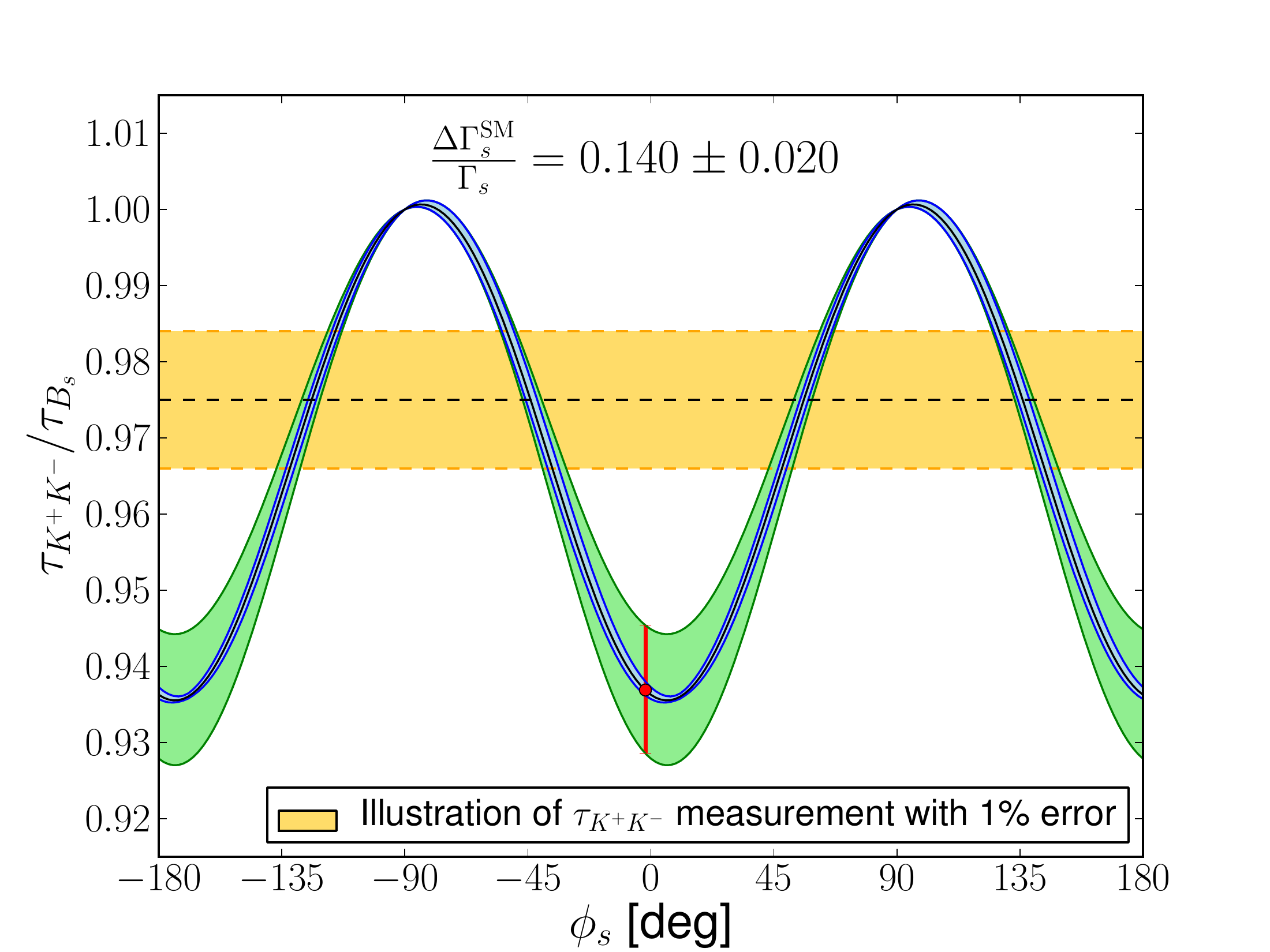} &
   \includegraphics[width=7.0truecm]{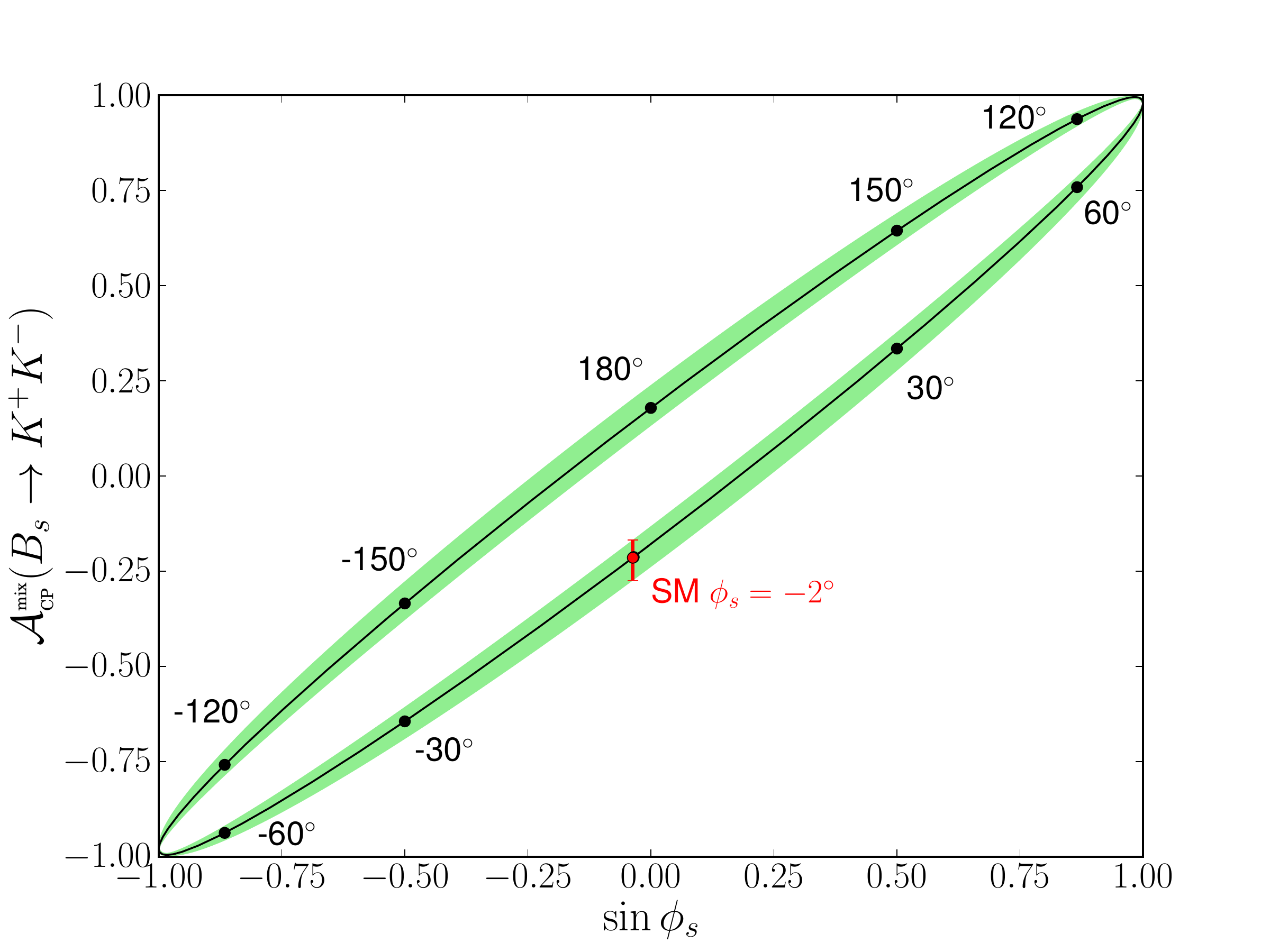}
   \end{tabular}}
      \caption{Dependence of  $\tau_{K^+K^-}$ and  ${\cal A}_{\rm CP}^{\rm mix}(B_s\to K^+K^-)$
      on $\phi_s$, as obtained and discussed
      in Ref.~[10].}\label{fig:1}
\end{figure}

Consequently, large CP-violating NP contributions to the decay amplitudes in Eq.~(\ref{ampl})
are already excluded. However, NP may well enter through $B^0_s$--$\bar B^0_s$ mixing. 
A particularly nice and simple observable is the effective $B^0_s\to K^+K^-$ lifetime 
$\tau_{K^+ K^-}$.\cite{FK} In the left panel of Fig.~\ref{fig:1}, the SM prediction for 
$\tau_{K^+ K^-}$ and its dependence on $\phi_s$ is shown. The major source of the theoretical 
uncertainty is the SM 
value of the width difference of the $B_s$ system whereas the uncertainties of the input parameters 
and $U$-spin-breaking corrections have a much smaller impact, as illustrated by the narrow band. 
LHCb has recently reported the first measurement of $\tau_{K^+ K^-}$,\cite{LHCb-tau} 
and the uncertainty
should soon be reduced significantly, where an error as illustrated in the figure appears
to be achievable. The next observable to enter the stage is the mixing-induced CP asymmetry 
${\cal A}_{\rm CP}^{\rm mix}(B_s\to K^+K^-)$. In the right panel of Fig.~\ref{fig:1}, 
its correlation with $\sin\phi_s$ is shown. The analysis of this observable turns out to be again
very robust with respect to the uncertainties of the input quantities. 

It becomes obvious that $\tau_{K^+ K^-}$ and ${\cal A}_{\rm CP}^{\rm mix}(B_s\to K^+K^-)$ offer
interesting probes for CP-violating NP in $B^0_s$--$\bar B^0_s$ mixing,\cite{FK} thereby
complementing the analyses of $B^0_s\to J/\psi\phi$. Once the CP asymmetries of the 
$B^0_s\to K^+K^-$, $B^0_d\to\pi^+\pi^-$ system have been measured by LHCb, $\gamma$ 
can be extracted in an optimal way.\cite{RF-BsK+K-} Based on the picture emerging from the 
current data, a stable situation with respect to $U$-spin-breaking corrections is expected.\cite{FK}

\subsection{$B^0_s\to\mu^+\mu^-$}
The $B^0_s\to\mu^+\mu^-$ channel originates from penguin and box topologies in 
the SM and is a well-known probe for NP. The upper bounds from the Tevatron and 
the first LHCb constraint, which was reported at Moriond 2011,\cite{LHCb-Bsmumu} 
are about one order of magnitude above the SM expectation of 
$\mbox{BR}(B_s\to\mu^+\mu^-)=(3.6\pm0.4)\times10^{-9}$, where the error is 
dominated by a lattice QCD input.\cite{buras}

Concerning the measurement of $B^0_s\to \mu^+\mu^-$ at LHCb, the major source
of uncertainty for the normalization of the branching ratio is $f_d/f_s$, where $f_q$
is the fragmentation function describing the probability that a $b$ quark hadronizes
as a $B_q$ meson.\cite{LHCb-roadmap}

\begin{figure}[!t]
  \centering
  \begin{tabular}{cc}
    \includegraphics[width=0.32\textwidth]{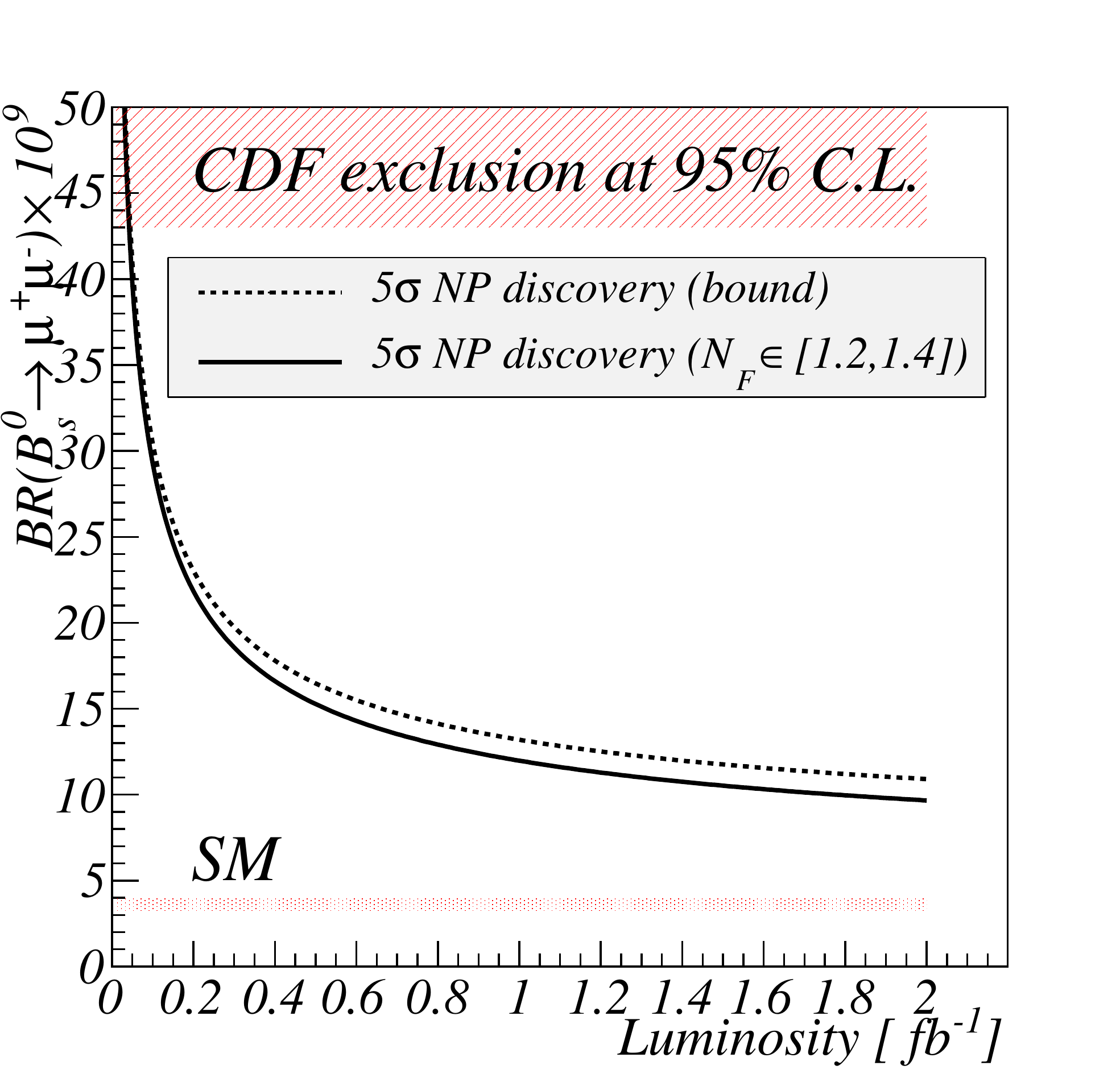} 
    \includegraphics[width=0.32\textwidth]{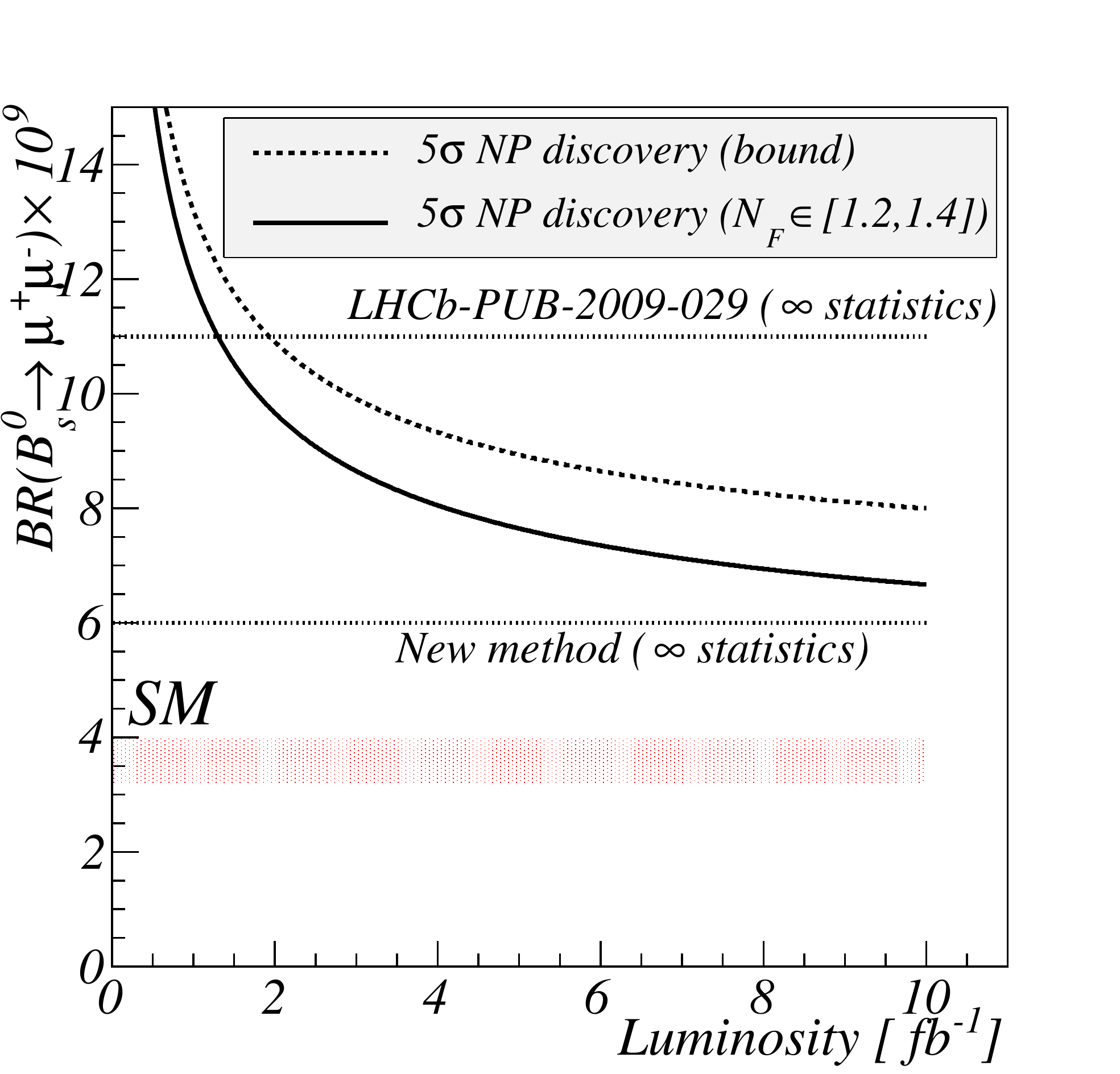} 
  \end{tabular} 
  \caption{Illustration of the LHCb NP reach in $B^0_s\to\mu^+\mu^-$ resulting
  from the strategy proposed in Ref.~[13].}\label{fig:2}
\end{figure}

In view of this challenge, a new strategy was proposed,\cite{FST-1} allowing the
measurement of $f_d/f_s$ at LHCb. The starting point is 
\begin{equation}
\frac{N_s}{N_d} = \frac{f_s}{f_d}\times \frac{\epsilon(B_s\rightarrow
  X_{1})}{\epsilon(B_d\rightarrow X_{2})}\times \frac{\mbox{BR}(B_s\rightarrow 
  X_{1})}{\mbox{BR}(B_d\rightarrow X_{2})},
\end{equation}
where the $N$ factors denote the observed number of events and the $\epsilon$ factors
are total detector efficiencies. Knowing the ratio of the branching ratios, $f_d/f_s$ could
be extracted. In order to implement this feature in practice, the $B_s\to X_1$ and $B_d\to X_2$
decays have to satisfy the following requirements: the ratio of their branching ratios must be 
``easy" to measure at LHCb, the decays must be robust with respect to the impact of NP, and 
the ratio of their BRs must be theoretically well understood. These requirements are
satisfied by the $U$-spin-related $\bar B^0_s\to D_s^+\pi^-$ and $\bar B^0_d\to D^+K^-$ decays.
In these channels, ``factorization" of hadronic matrix elements is expected to work very well,
and the theoretical precision is limited by non-factorizable $U$-spin-breaking effects, leading
to an uncertainty at the few-percent level. These features can also be tested through experimental 
data, supporting this picture.\cite{FST-2} The NP discovery potential in $B^0_s\to\mu^+\mu^-$ 
at LHCb resulting from this method is illustrated in Fig.~\ref{fig:2},
which shows the smallest value of $\mbox{BR}(B^0_s\to\mu^+\mu^-)$ allowing the detection 
of a $5 \sigma$ deviation from the SM as a function of the luminosity at LHCb 
(at the nominal beam energy of 14 TeV).\cite{FST-1} 

LHCb has reported the first results for $f_s/f_d$ from this strategy at this conference,\cite{LHCb-fsfd} 
yielding $f_s/f_d= 0.245 \pm 0.017|_{\rm stat} \pm 0.018|_{\rm sys}  \pm 0.018|_{\rm theo}$. This is 
an average over the data for the $\bar B^0_s\to D_s^+\pi^-$, $\bar B^0_d\to D^+K^-$ and $\bar 
B^0_s\to D_s^+\pi^-$, $\bar B^0_d\to D^+\pi^-$ systems, where the latter
offers a variant of the method for extracting $f_s/f_d$.\cite{FST-2}

\section{Concluding Remarks}
We are moving towards new frontiers in $B$ physics. There are good chances that these
studies will reveal first footprints of NP at the LHC. Exciting years are ahead of us!

%\section*{Acknowledgments}
%This is where one places acknowledgments for funding bodies etc.
%Note that there are no section numbers for the Acknowledgments, Appendix
%or References.


\section*{References}
\begin{thebibliography}{99}
\bibitem{buras} A.~J.~Buras,
  %``Flavour Theory: 2009,''
  PoS E {\bf PS-HEP2009} (2009) 024
  [arXiv:0910.1032 [hep-ph]].
  %%CITATION = POSCI,EPS-HEP2009,024;%%

  \bibitem{DDF}A.~S.~Dighe, I.~Dunietz and R.~Fleischer,
  %``Extracting CKM phases and $B_s - \bar{B}_s$ mixing parameters from angular
  %distributions of nonleptonic $B$ decays,''
  Eur.\ Phys.\ J.\  C {\bf 6} (1999) 647.
 % [arXiv:hep-ph/9804253].
  %%CITATION = EPHJA,C6,647;%%

\bibitem{DFN}I.~Dunietz, R.~Fleischer and U.~Nierste,
  %``In pursuit of new physics with $B_s$ decays,''
  Phys.\ Rev.\  D {\bf 63} (2001) 114015.
  %[arXiv:hep-ph/0012219].
  %%CITATION = PHRVA,D63,114015;%%
  
\bibitem{LHCb-phis}LHCb Collaboration, LHCb-CONF-2011-002.
  
\bibitem{FFM}S.~Faller, R.~Fleischer and T.~Mannel,
  %``Precision Physics with $B^0_s \to J/\psi \phi$ at the LHC: The Quest for
  %New Physics,''
  {\em Phys.\ Rev.}\  D {\bf 79} (2009) 014005.
 % [arXiv:0810.4248 [hep-ph]].
  %%CITATION = PHRVA,D79,014005;%%

\bibitem{RF-BspsiKS}R.~Fleischer,
  %``Extracting $\gamma$ from $B(s/d) \to J/\psi K_{S}$ and $B(d/s) \to D^+(d/s)
  %D^-(d/s)$,''
  {\em Eur.\ Phys.\ J.}\  C {\bf 10} (1999) 299.
 % [arXiv:hep-ph/9903455].
  %%CITATION = EPHJA,C10,299;%%

\bibitem{DeBFK} K.~De Bruyn, R.~Fleischer and P.~Koppenburg,
  %``Extracting gamma and Penguin Topologies through CP Violation in B_s^0 ->
  %J/psi K_S,''
  {\em Eur.\ Phys.\ J.}\  C {\bf 70} (2010) 1025.
 % [arXiv:1010.0089 [hep-ph]].
  %%CITATION = EPHJA,C70,1025;%%
  
\bibitem{RF-BsK+K-}R.~Fleischer,
  %``New strategies to extract Beta and gamma from B(d) ---> pi+ pi- and B(S)
  %---> K+ K-,''
  {\em Phys.\ Lett.}\  B {\bf 459} (1999) 306;
%  [arXiv:hep-ph/9903456].
  %%CITATION = PHLTA,B459,306;%%
 % R.~Fleischer,
  %``$B_{s,d} \to \pi \pi, \pi K, KK$: Status and Prospects,''
  {\em Eur.\ Phys.\ J.}\  C {\bf 52} (2007) 267.
%  [arXiv:0705.1121 [hep-ph]].
  %%CITATION = EPHJA,C52,267;%%

\bibitem{LHCb-roadmap}B. Adeva {\it et al.}\  [LHCb Collaboration], LHCb-PUB-2009-029
  %``Roadmap for selected key measurements of LHCb,''
 [arXiv:0912.4179 [hep-ex]].
  %%CITATION = ARXIV:0912.4179;%%

\bibitem{FK} R.~Fleischer and R.~Knegjens,
  %``In Pursuit of New Physics with B^0_s -> K^+K^-,''
  {\em Eur.\ Phys.\ J.}\  C {\bf 71} (2011) 1532.
%  [arXiv:1011.1096 [hep-ph]].
  %%CITATION = EPHJA,C71,1532;%%
  
\bibitem{LHCb-tau}LHCb Collaboration, LHCb-CONF-2011-018.

\bibitem{LHCb-Bsmumu}R.~Aaij {\it et al.}  [LHCb Collaboration],
  %``Search for the rare decays Bs -->mumu and Bd -->mumu,''
  {\em Phys.\ Lett.}\  B {\bf 699} (2011) 330.
 % [arXiv:1103.2465 [hep-ex]].
  %%CITATION = PHLTA,B699,330;%%

\bibitem{FST-1}R.~Fleischer, N.~Serra and N.~Tuning,
  %``A New Strategy for B_s Branching Ratio Measurements and the Search for New
  %Physics in B^0_s -> mu^+ mu^-,''
  {\em Phys.\ Rev.}\  D {\bf 82} (2010) 034038.
 % [arXiv:1004.3982 [hep-ph]].
  %%CITATION = PHRVA,D82,034038;%%

\bibitem{FST-2}R.~Fleischer, N.~Serra and N.~Tuning,
  %``Tests of Factorization and SU(3) Relations in B Decays into Heavy-Light
  %Final States,''
  {\em Phys.\ Rev.}\  D {\bf 83} (2011) 014017.
 %  [arXiv:1012.2784 [hep-ph]].
  %%CITATION = PHRVA,D83,014017;%%
  
\bibitem{LHCb-fsfd}LHCb Collaboration, LHCb-CONF-2011-013.
  
  \end{thebibliography}
\end{document}